\documentclass[a4paper,fleqn]{cas-dc}
\usepackage[justification=centering]{caption}
\usepackage{bm}
\usepackage[justification=centering]{caption}
\usepackage[numbers]{natbib}
\usepackage{graphicx}
\usepackage{multirow} 
\usepackage{stfloats} 
\usepackage[export]{adjustbox}
\usepackage{subfig}
\usepackage{bbding}
\usepackage{amsmath}
\usepackage{amssymb}
\usepackage{booktabs}
\usepackage{hyperref}
\usepackage[misc]{ifsym}
\usepackage{color}
\usepackage{caption}
\usepackage[normalem]{ulem} 
\usepackage{xcolor}

\hypersetup{
    colorlinks=false,
    allbordercolors=red,
    pdfborderstyle={/S/U/W 1}
}

\def\tsc#1{\csdef{#1}{\textsc{\lowercase{#1}}\xspace}}
\tsc{WGM}
\tsc{QE}
\tsc{EP}
\tsc{PMS}
\tsc{BEC}
\tsc{DE}

\begin{document}
\let\WriteBookmarks\relax
\def\floatpagepagefraction{1}
\def\textpagefraction{.001}
\shorttitle{\title{HEC-GCN: Hypergraph Enhanced Cascading Graph Convolution Network for Multi-Behavior Recommendation}}
\shortauthors{Yin et~al.}


\title[mode = title]{HEC-GCN: Hypergraph Enhanced Cascading Graph Convolution Network for Multi-Behavior Recommendation}

\author[1]{Yabo Yin}
\ead{yinyabo@stu.cqut.edu.cn}
\credit{Conceptualization of this study, Methodology, Software, Writing - original draft}
\address[1]{College of Computer Science and Engineering,  Chongqing University of Technology, Chongqing 400054, China}

\author[1]{Xiaofei Zhu}
\ead{zxf@cqut.edu.cn}
\credit{Conceptualization of this study, Methodology, Writing - original draft \& review \& editing, Supervision}
\cormark[1] 

\author[2]{Wenshan Wang}
\ead{wangwenshan@ict.ac.cn}
\credit{Methodology, Supervision}
\address[2]{Institute of Computing Technology, Chinese Academy of Sciences, Beijing 100190, China}

\author[3]{Yihao Zhang}
\ead{yhzhang@cqut.edu.cn}
\credit{Supervision}
\address[3]{School of Artificial Intelligence, Chongqing University of Technology, Chongqing, 400054, China} 

\author[4]{Pengfei Wang}
\ead{wangpengfei@bupt.edu.cn}
\credit{Supervision}
\address[4]{School of Computer Science, Beijing University of Posts and Telecommunications, Beijing 100876, China}


\author[2]{Yixing Fan}
\ead{fanyixing@ict.ac.cn}
\credit{Supervision}

\author[2]{Jiafeng Guo}
\ead{guojiafeng@ict.ac.cn}
\credit{Supervision}

\cortext[cor1]{Corresponding author}




\begin{abstract}
Multi-behavior recommendation (MBR) has garnered growing attention recently due to its ability to mitigate the sparsity issue by inferring user preferences from various auxiliary behaviors to improve predictions for the target behavior. 
Although existing research on MBR has yielded impressive results, they still face two major limitations. First, previous methods mainly focus on modeling fine-grained interaction information between users and items under each behavior, which may suffer from sparsity issue. Second, existing models usually concentrate on exploiting dependencies between two consecutive behaviors, leaving intra- and inter-behavior consistency largely unexplored.
To the end, we propose a novel approach named \underline{H}ypergraph \underline{E}nhanced \underline{C}ascading \underline{G}raph \underline{C}onvolution \underline{N}etwork for multi-behavior recommendation (HEC-GCN). To be specific, we first explore both fine- and coarse-grained correlations among users or items of each behavior by simultaneously modeling the behavior-specific interaction graph and its corresponding hypergraph in a cascaded manner. Then,  we propose a behavior consistency-guided alignment strategy that ensures consistent representations between the interaction graph and its associated hypergraph for each behavior, while also maintaining representation consistency across different behaviors.
Extensive experiments and analyses on three public benchmark datasets demonstrate that our proposed approach is consistently superior to previous state-of-the-art methods due to its capability to effectively attenuate the sparsity issue as well as preserve both intra- and inter-behavior consistencies. 
The code is available at https://github.com/marqu22/HEC-GCN.git.

\end{abstract}


\maketitle
\section{Introduction} 
Multi-behavior recommendation (MBR) \cite{gao2021learning,jin2020multi,yan2023cascading-crgcn,cheng2023multi},  has emerged as an important research problem in recent years and garnered significant attention across various communities. 
Compared to traditional recommendation methods which solely rely  on a single type of behavior, MBR makes recommendations by leveraging multiple user-item interaction information in order to alleviate the data sparsity issue.  Besides the  target behavior (e.g., buy), a wide range of auxiliary behaviors (e.g., view, click and cart) can be  utilized to boost recommendation performance.

Early studies \cite{tang2016empirical,zhao2015improving}  can be regarded  as straightforward extensions of single-behavior recommendation models, aiming to obtain better user and item representations in the target behavior by utilizing shared embeddings across various behaviors.
Meanwhile, some findings \cite{loni2016bayesian,ding2018implicit} also reveal that certain sampling strategies can boost model recommendation performance. 
Later, many research efforts have been devoted to leveraging Deep Neural Networks (DNN) for the task of MBR due to its strong capability to capture the complex relationships between users and items, and have achieved encouraging performance. These include methods that employ hierarchical attention mechanism to model inner parts within a behavior as well as inter-view relations between different behaviors \cite{guo2019buying}, or transfer the prediction of low-level behavoirs to high-level behaviors \cite{chen2020efficient}.

Recently, a number of graph convolutional networks (GCN) based methods have been proposed to explicitly model the high-order dependencies between different types of user-item interaction data via utilizing the graph-based message passing mechanism. For example, 
MBGCN \cite{jin2020multi} leverages multi-behavior data to construct a unified graph, and applies both user-item propagation and item-item propagation to learn behavior strength and behavior semantics, respectively. 
GNMR \cite{xia2021multi} builds a multi-behavior interaction graph and performs embedding propagation for capturing complex relationships among different types of user behaviors. 
S-MBRec \cite{gu2022self} introduces supervised signal to consider the differences of various behaviors and develops a star-style contrastive learning to model the commonality between auxiliary and target behaviors. 
MB-HGCN \cite{yan2023mbhgcn} learns behavior-specific and global  embeddings based on each individual behavior graph and a unified homogeneous graph. The multi-task learning strategy is leveraged for joint optimization. 
BCIPM \cite{yan2024behavior}  considers the variation of user preferences across different behaviors and distills item-aware preferences from user-item interactions. 
Some research works  \cite{gao2019neural,yan2022cascading,cheng2023multi,meng2023parallel} propose to exploit the cascading relationship among different behaviors to enhance the performance of recommendation, where useful information learned from preceding behaviors will be utilized for refining the learned embeddings of successive behaviors.

Despite the encouraging progress obtained by conventional methods, there are still two key issues limiting their efficacy.
Firstly, due to the sparsity of user interaction data and the limitation of the bipartite graph structure of user-item relationships, employing multi-layer graph message propagation can easily lead to the over-smoothing problem \cite{min2020scattering,zhou2020towards}, resulting in model performance degradation.  
Existing research works mainly rely on modeling fine-grained interactions between users and items under specific behaviors, while they overlook to explore the coarse-grained  information existing in the interaction graph, e.g., certain users may have similar interest preferences, or specific items may belong to the same category. Since the coarse-grained information can effectively complement the fine-grained interaction information, leveraging both simultaneously is valuable for capturing more accurate user preferences.
Secondly, previous cascading based models \cite{cheng2023multi, yan2022cascading} emphasize exploiting dependencies  in  the sequence of behaviors for embedding learning, e.g., the embeddings learned from one behavior will be fed into the next behavior's embedding learning process. Although they can effectively capture the dependency among two consecutive behaviors, the rich global consistency among different behaviors  as well as the intra-behavior dependency signals are largely ignored.

In order to address the issues mentioned above, in this paper, we propose a novel approach named \underline{H}ypergraph \underline{E}nhanced \underline{C}ascading \underline{G}raph \underline{C}onvolution \underline{N}etwork for multi-behavior recommendation (HEC-GCN).
Specifically, we first propose a hypergraph enhanced cascading graph convolution network by introducing a user-item interaction graph and a hypergraph for each  behavior. 
The behavior-specific hypergraph is developed to capture coarse-grained correlations among users or items, which can alleviate the sparsity issue within the user-item interaction graph. 
Then, we develop a behavior consistency-guided contrastive learning module, which consists of an intra-behavior consistency contrastive learning submodule and a cross-behavior consistency contrastive learning submodule. The former aims to align user and item embeddings inferred from an interaction graph to their counterparts learned from the corresponding  hypergraph. The latter attempts to align embeddings of users and items from each behavior towards the common embeddings derived from  a global graph with user-item interactions from all behaviors.
We conduct extensive experiments on three real-world benchmark datasets, i.e., Beibei, Taobao and Tmall, to verify the effectiveness of our proposed approach. Experimental results demonstrate that HEC-GCN outperforms all competitive baselines with a large margin, e.g., the relative performance improvements of HEC-GCN over the best performing baseline on the three datasets (i.e., Beibei, Taobao and Tmall) are 19.20\%, 37.45\% and 13.43\% in terms of HR@10, respectively.

To summarize, the main contributions are as follows:
\begin{itemize}
    \item We introduce a novel hypergraph enhanced cascading graph convolution network, which explores both fine- and coarse-grained correlations among users or items  by simultaneously modeling the user-item interaction graph and its corresponding hypergraph for each behavior. 
    \item We propose a behavior consistency-guided alignment strategy to maintain both intra- and inter-behavior consistency based on contrastive learning. 
    \item We conduct extensive experiments on three real-world datasets to examine the effectiveness of HEC-GCN. Experimental results present a considerable performance improvement compared to the state-of-the-art baseline methods.
\end{itemize}

\section{Related Work}
Multi-behavior recommendation aims to utilize valuable signals from auxiliary behaviors (e.g., view, click, cart) to  mitigate the sparsity issue in the target behavior (e.g., buy). 
Early methods \cite{singh2008relational,tang2016empirical,zhao2015improving} mainly rely on  extending the single-behavior recommendation technique such as matrix factorization by introducing shared embeddings across different behaviors. 
Some works \cite{ding2018implicit,guo2017resolving,loni2016bayesian} apply sampling strategies to incorporate information from auxiliary behaviors into the target behavior, which are tailored to inherent characteristics in multi-behavior scenarios.

With the great success of deep neural network (DNN) in recommendation systems, many DNN-based methods have been proposed to model user preferences  via capturing  complex relationships  between users and items \cite{huang2021recent}. 
For example, 
DIPN \cite{guo2019buying} develops a hierarchical attention mechanism which consists of two attention layers, where a bottom attention layer is designed to model  inner parts of each behavior and a top attention layer is utilized to learn the inter-view relations between different behaviors.
EHCF \cite{chen2020efficient} attempts to transfer the prediction of low-level behaviors to high-level behaviors to capture the complicated relationships among different behaviors.
ARGO \cite{wu2021argo} captures the characteristics of each user by developing a neural collaborative filtering model and  models the ordinal relation among multiple behaviors by correlating each behavior's prediction.

Recently, graph convolutional network (GCN) has attracted lots of attention for multi-behavior recommendation \cite{xia2021knowledge,guo2023compressed,meng2023hierarchical,meng2024coarse}, which represents the interaction relationships between users and items as graph structure and adopts the message propagation paradigm to learn  node embeddings. For instance, 
MBGCN \cite{jin2020multi} represents multi-behavior information by constructing a unified graph and develops a multi-behavior graph convolutional network. It employs a user-item propagation layer and an item-item propagation layer to capture the strength and semantics of behaviors, respectively.
GNMR \cite{xia2021multi} captures the implicit dependencies among different behaviors via a relation dependency encoder, and models the graph-structured interactions by conducting embedding propagation over a multi-behavior interaction graph.  
S-MBRec \cite{gu2022self} employs GCN to learn user preferences. Specifically, It applies a supervised task to model the differences between embeddings, and utilizes a star-style contrastive learning task to model embedding commonality between auxiliary and target behaviors.
MB-HGCN \cite{yan2023mbhgcn} employs a hierarchical graph convolutional network which consists of global embedding learning and behavior-specific embedding learning.  
Since directly transferring information from auxiliary behaviors to the target behavior may inject noise due to the variation of user preferences across different behaviors. 
BCIPM \cite{yan2024behavior} proposes to extract item-aware preferences from user-item interaction data of each behavior and only consider item-aware preferences related to the target behavior for recommendation. 

As there are usually certain orders between different behaviors, some research works concentrate on capturing such behavior dependencies to learn better user preferences. 
NMTR \cite{gao2019neural} learns user preferences by assuming that there is a shared embedding across different behaviors, and explores the ordinal relationship information  by correlating the model prediction in a cascaded manner. 
CRGCN \cite{yan2022cascading} proposes a cascading residual graph convolutional network to model user preferences by exploiting the connections between consecutive behaviors, and adopts the multi-task learning to conduct jointly optimization by comprehensively exploiting supervision signals from different behaviors.
MB-CGCN \cite{cheng2023multi} explicitly models the behavior dependencies for learning user preferences,  which are aggregated for the target behavior prediction. 
PKEF \cite{meng2023parallel}  extends the cascade paradigm by incorporating parallel knowledge to learn better representations for different behaviors  while learning hierarchical correlation information to alleviate the issue caused by the imbalanced behavior distribution. 

Unlike studies mentioned above, our method incorporates a behavior-specific hypergraph  to capture coarse-grained correlations among users or items for each behavior, and simultaneously leverage both fine- and coarse-grained information to comprehensively model user preferences. Moveover, we develop  a behavior consistency-guided contrastive learning to maintain both intra- and inter-behavior consistency for each behavior.

\section{Problem Formulation}
In this work, we focus on developing a novel recommendation model by exploiting  auxiliary behaviors to enhance the learning of  better user preferences in the target behavior. 
We describe the basic notations frequently used in this paper. 
Let $\mathcal{U} = \{u_1,u_2,...,u_M\}$ and $\mathcal{I} = \{i_1, i_2, ..., i_N\}$ denote the set of users and items, where $M$ and $N$ denote the number of users and items, respectively. 
We use  $k$ ($1 \leq k \leq K$) to denote the $k$-th behavior, where $K$ denotes the number of users' behaviors. 
$\mathcal{G}_k$ is the user-item interaction graph for the $k$-th behavior, and $\mathcal{R}^k$ is the interaction matrix of the corresponding behavior.  For $r^k_{ui} \in \mathcal{R}^k$, $r^k_{ui}=1$ indicates that there is an interaction between   user $u$ and   item $i$ under the $k$-th behavior, otherwise $r^k_{ui}=0$.
In addition, we define a global graph $\mathcal{G}_g=\cup_{k=1}^{K}\mathcal{G}_k $ and its corresponding interaction matrix is  $\mathcal{R}^g$. 
For $r^g_{ui} \in \mathcal{R}^g$, $r^g_{ui}=1$ means that   user $u$ and   item $i$ have interacted at least once in these $K$ behaviors. 
The MBR task is then formulated as follows:

\textbf{Input}:  A user set $\mathcal{U}$, an item set $\mathcal{I}$, a list of behavior interaction graphs  $[\mathcal{G}_1, \mathcal{G}_2, ..., \mathcal{G}_K]$, and  a global graph $\mathcal{G}_g$.

\textbf{Output}: A  predicted score, which indicates the probability that user $u$ will interact with item $i$ in the $K$-th behavior (i.e., target behavior).

\section{Proposed Method}
We propose a Hypergraph Enhanced Cascading Graph Convolution Network (HEC-GCN) for multi-behavior recommendation, which is  comprised of four modules, i.e.,
\textbf{(1)}  A \textit{Global Graph Learning}  module, which is dedicated to acquiring insights into users' general preference across distinct behaviors.
\textbf{(2)}  A \textit{Hypergraph Enhanced Cascading Graph Convolution Network}, which follows the cascading paradigm and incorporates  behavior hypergraphs to enhance the learning of representations under specific behaviors.
\textbf{(3)} A \textit{Behavior Consistency-Guided Contrastive Learning}  module encompassing both intra- and inter-behavior intrinsic consistency. 
\textbf{(4)} A \textit{Prediction} module that employs the multi-task learning to conduct joint optimization. 

\begin{figure*}[!t]
\centering
\includegraphics[width=\textwidth]{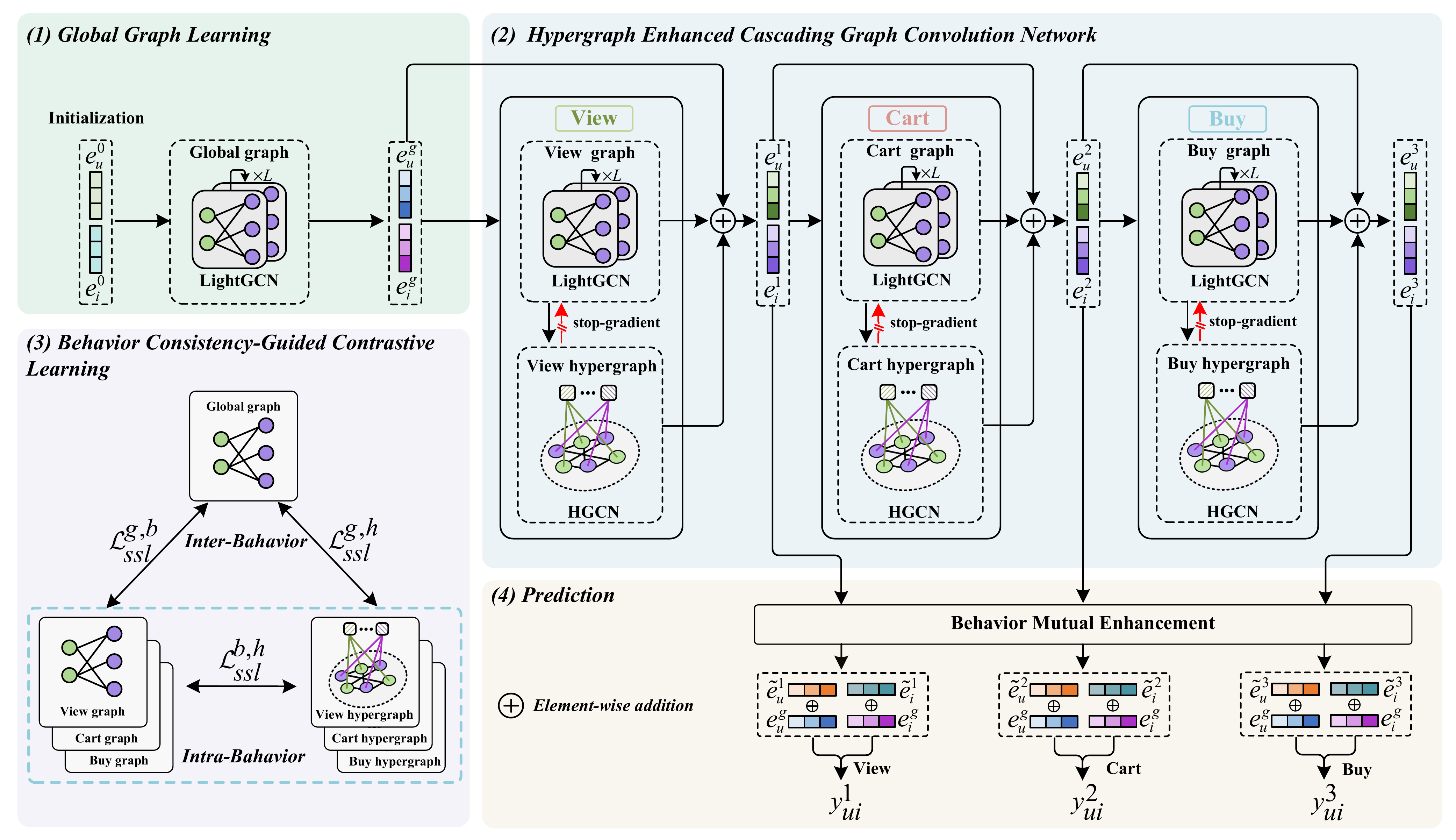}
\caption{
The overall architecture of HEC-GCN.
We utilize three behaviors (i.e., {view, cart and buy})  as an example, wherein buy is the target behavior.
}
\label{fig:model_overall}
\end{figure*}

\subsection{ Global Graph Learning}
We introduce the global  graph $\mathcal{G}_g$ to serve as an indicator of whether user $u$ and item $i$ have interacted at least once across all behaviors, i.e., $\mathcal{G}_g$ contains all user interactions across multiple behaviors.  We utilize the Xavier \cite{xavier2010understanding} to initialize user $u$ and item $i$ as $d$-dimensional embeddings, denoted as    ${e}^{0}_{u} \in \mathbb{R}^{d}$, ${e}^{0}_{i} \in \mathbb{R}^{d}$, respectively.  Then, we update the embeddings of both user and item by utilizing LightGCN \cite{he2020lightgcn}: 
\begin{equation}
\begin{split}
e_{u}^{(l)}  &= \sum_{i \in \mathcal{N}_{u}} \frac{1}{\sqrt{\left|\mathcal{N}_{u}\right|} \sqrt{\left|\mathcal{N}_{i}\right|}} e_{i}^{(l-1)}, \\
e_{i}^{(l)}  &= \sum_{u \in \mathcal{N}_{i}} \frac{1}{\sqrt{\left|\mathcal{N}_{i}\right|} \sqrt{\left|\mathcal{N}_{u}\right|}} e_{u}^{(l-1)},
\end{split}
\label{eq:lightgcn}
\end{equation} 
where $\mathcal{N}_u$ denotes the set of items that user $u$ has interacted with, and $\mathcal{N}_i$ denotes the set of users who have interacted with item $i$. 
$e_{u}^{(l)}$ and $e_{i}^{(l)}$  denote  the embeddings of user $u$ and item $i$ after the graph  convolution of the $l$-layer, respectively. 
$e_{u}^{(0)}$ and $e_{i}^{(0)}$ are initialized with ${e}^{0}_{u}$ and ${e}^{0}_{i}$, respectively. 

We obtain the global embedding representations of user $u$ and item $i$ by aggregating the outputs of all layers: 
\begin{align}
e_{u}^{g}  = \sum_{l  =  0}^{L} e_u^{(l)}, \\
e_{i}^{g}  = \sum_{l  =  0}^{L} e_i^{(l)},
\end{align} 
where $L$ is the total number of graph convolution layers.  

\subsection{Hypergraph Enhanced Cascading Graph Convolution Network}
We propose a hypergraph enhanced cascading graph convolution network to learn user and item representations from the user-item interaction graph and its corresponding hypergraph for each behanvior in a cascaded manner, where the learned representations from a preceding behavior will be fed into its subsequent behavior. Based on this, we can capture the higher-order collaborative relationships among users and items, thereby enhancing the model capability of learning better user preferences.   

\subsubsection{\textbf{Behavior-Specific Interaction  Graph Learning.}}
For the $k$-th behavior, we  employ LightGCN on its corresponding user-item interaction graph $\mathcal{G}_k$  to learn representation of each node. Specifically, we update node representations by aggregating information from their neighbors as follows: 
\begin{align}
\textbf{e}_u^{k,(l)} = \sum_{i\in \mathcal{N}_u} \frac{1}{\sqrt{|\mathcal{N}_u|}\sqrt{|\mathcal{N}_i|}} \textbf{e}_i^{k,(l-1)},\\
\textbf{e}_i^{k,(l)} = \sum_{u\in \mathcal{N}_i} \frac{1}{\sqrt{|\mathcal{N}_i|}\sqrt{|\mathcal{N}_u|}} \textbf{e}_u^{k,(l-1)},
\end{align}
where $\textbf{e}_u^{k,(l)}$ and $\textbf{e}_i^{k,(l)}$ respectively denote the updated representations of user $u$ and item $i$ after $l$ layers propagation on the $k$-th interaction graph.
Note that for the $k$-th behavior-specific interaction graph, we initialize its embeddings of users and items with the corresponding embeddings learned from its preceding behavior: 
\begin{align}
e_{u}^{k,(0)}  = e_{u}^{k-1}, \\
e_{i}^{k,(0)}  = e_{i}^{k-1}, 
\end{align}
where $e_{u}^{k-1}$ and $ e_{i}^{k-1}$ denote the learned embeddings of users and items from its preceding behavior-specific interaction graph. Meanwhile, we initialize the embeddings for the first behavior by utilizing the user and item embeddings obtained from global graph learning.
After that, we obtain the embeddings of   user $u$ and   item $i$ of the $k$-th behavior as follows: 
\begin{align}
e_{u}^{b,k}  = \sum_{l  =  0}^{L} e_u^{k,(l)}, \\
e_{i}^{b,k}  = \sum_{l  =  0}^{L} e_i^{k,(l)}.
\end{align}

\subsubsection{\textbf{Behavior-Specific Hypergraph Learning.}}
To alleviate the impact of noisy interactions and model the complex high-order correlations among users or items, we augment our approach by introducing additional learnable behavior-specific hypergraphs. Differ from conventional graph structure,  hypergraph allows a edge to connect to more than two nodes \cite{fan2022heterogeneous}, offering a  comprehensive representation of the intricate interactions inherent in each behavior.  

Specifically, we first define a set of hyperedge representations $\mathcal{H}^k_U \in \mathbb{R}^{M \times S}$ for all $M$ users on the $k$-th behavior, where 
$S$ denotes the number of hyperedges. Similarly, the hyperedge representations for all $N$ items are defined as $\mathcal{H}^k_I \in \mathbb{R}^{N \times S}$.
Inspired by the prior work \cite{xia2022hypergraph}, we address computational efficiency and mitigate over-fitting concerns within the hypergraph learning module. Specifically, we employ parameterized hyperedges, which facilitate hyperedge parameter optimization in a low-rank manner. This operation is formulated as:
\begin{align}
\mathcal{H}^k_U  = E^{b,k}_{U} \cdot W^{h,k}_{U}, \\
\mathcal{H}^k_I  = E^{b,k}_{I} \cdot W^{h,k}_{I},
\end{align}
where $ W^{h,k}_{U} \in \mathbb{R}^{d \times S}$ and  $W^{h,k}_{I} \in \mathbb{R}^{d \times S}$ are the learnable parameters, 
$E_U^{b,k}=[e_{u_1}^{b,k},...,e_{u_M}^{b,k}] \in \mathbb{R}^{M \times d} $ and  $E_I^{b,k}=[e_{i_1}^{b,k},...,e_{i_N}^{b,k}]  \in \mathbb{R}^{N \times d}$ represent  the embeddings  of all users and items corresponding to the $k$-th behavior-specific interaction graph, respectively. 
For hypergraph learning  in the $k$-th behavior, we  utilize a simplified  hypergraph  convolution network (HGCN) \cite{wang2024hgrec}   for the user-side and item-side hypergraphs of each behavior:
\begin{align}
E_U^{h,k} = (\mathcal{H}^k_U \cdot {(\mathcal{H}^k_U)}^{\mathsf{T}}) \cdot E_U^{b,k}, \\
E_I^{h,k} = (\mathcal{H}^k_I \cdot {(\mathcal{H}^k_I)}^{\mathsf{T}}) \cdot E_I^{b,k}, 
\end{align} 
where $(\mathcal{H}^k_U \cdot {(\mathcal{H}^k_U)}^{\mathsf{T}}) \in \mathbb{R}^{M \times M}$ and  $(\mathcal{H}^k_I \cdot {(\mathcal{H}^k_I)}^{\mathsf{T}}) \in \mathbb{R}^{N \times N}$ denote the  hypergraph dependency matrices for the user-side and the item-side, respectively.  

After hypergraph learning in the $k$-th behavior, the  hypergraph embeddings of user $u$ and item $i$ are represented as $e_{u}^{h,k} \in \mathbb{R}^{d}$ and $e_{i}^{h,k} \in \mathbb{R}^{d}$, respectively.
It is worth noting that the hypergraph learning developed in this work is different from  previous works \cite{xia2022hypergraph}. First, to incorporating more user interaction information during the hypergraph convolution, we inject the learned embeddings from the behavior interaction graph into the  process of hypergraph learning. Second, to preserve the independence of behavior interaction graph learning,  we adopt a stop-gradient strategy to ensure unidirectional information propagation from behavior-specific interaction graph learning to behavior-specific hypergraph learning. Third,  to simplify the process of information propagation in the hypergraph learning, we remove activation functions from the hypergraph convolution. 
 
\subsubsection{\textbf{Behavior-Specific Embedding Integration.}}
With the learned embeddings of users and items from the $k$-th behavior-specific interaction graph and behavior-specific hypergraph, we integrate them to obtain their corresponding behavior-specific embeddings:
\begin{align}
e_u^{k} = e_u^{b,k} \oplus e_u^{h,k} \oplus e_u^{k-1} , \\
e_i^{k} = e_i^{b,k} \oplus e_i^{h,k} \oplus e_i^{k-1},
\end{align}
where $e_u^{k-1}$ and $e_i^{k-1}$ are residual-connected embeddings derived from the previous behavior, $e_u^{k}$ and $e_i^{k}$ denote the updated embeddings for user $u$ and item $i$ from the $k$-th behavior, which will be utilized as initialization embeddings for the subsequent behavior.  
Note that we initialize the embeddings of user $u$ and item $i$  of the first behavior with the corresponding global embedding representations $e_u^g$ and $e_i^g$, respectively.

\subsection{Behavior Consistency-Guided Contrastive Learning}
Although the cascading architecture effectively captures dependencies between consecutive behaviors, it falls short in fully exploring users' shared interest preferences across different behaviors and their multi-granular interest preferences within behavior-specific interaction graphs and hypergraphs. To address this, we introduce inter- and intra-behavior consistency constraints through a contrastive learning framework, enabling a more comprehensive modeling of users' interest preferences from multiple perspectives. 
 
\subsubsection{{Inter-Bahavior Consistency Contrastive Learning.}} 
Given the variations in behavioral patterns, directly constraining the embedding consistency of users or items across different behaviors may hinder the exploration of common information across them. Consequently, we employ the global graph as an intermediary and align  embeddings of nodes  learned from each behavior graph 
towards that from the global graph.  
Specifically, we enhance the similarity of embeddings for the same node under both the global graph $\mathcal{G}_g$ and the $k$-th behavior graph $\mathcal{G}_k$ (i.e, positive sample pairs ($e^g_u,e^{b,k}_u$)), while the embeddings for different users are encouraged to diverge from each other (i.e, negative sample pairs ($e^g_u,e^{b,k}_{v}$)).  
Similar to InfoNCE \cite{oord2018representation}, we formalize the  inter-behavior  contrastive loss $\mathcal{L}^{g,b}_{cl_{},user}$ between the global graph and all $K$ behavior-specific interaction graphs on the user-side as follows: 
\begin{equation}
    \mathcal{L}^{g,b}_{cl_{},user}=\sum_{k=1}^{K}\sum_{u \in \mathcal{U}}-\log \frac{\exp (\phi(e^g_u, e^{b,k}_u) / \tau)}{\sum_{v \in \mathcal{U}} \exp (\phi(e^g_u, e^{b,k}_{v}) / \tau)}, 
\label{eq:G-B_loss}
\end{equation}    
where $\tau$ denotes the temperature coefficient, and $\phi(\cdot)$ is the similarity function.  
The computation of $\mathcal{L}^{g,b}_{cl,item}$ on the item-side follows a similar way. The comprehensive contrastive loss $\mathcal{L}^{g,b}_{cl}$ is obtained by combining the user-side and the item-side losses:
\begin{align}
    \mathcal{L}^{g,b}_{cl} = \mathcal{L}^{g,b}_{cl,user} +\mathcal{L}^{g,b}_{cl,item}.
\end{align} 

Similarly, we obtain the inter-behavior contrastive loss $\mathcal{L}^{g,h}_{cl}$ between the global graph and all $K$ behavior-specific hypergraphs as follows:
\begin{align}
    \mathcal{L}^{g,h}_{cl} = \mathcal{L}^{g,h}_{cl,user} +\mathcal{L}^{g,h}_{cl,item},
\end{align}
where $\mathcal{L}^{g,h}_{cl,user}$ and $\mathcal{L}^{g,h}_{cl,item}$ represent the contrastive losses on the user-side and item-side, respectively.


\subsubsection{\textbf{ Intra-Bahavior Consistency Contrastive Learning.}} 
To capture the intrinsic consistency between each user-item interaction graph and its corresponding hypergraph under the same behavior, we design the intra-behavior  contrastive learning to maintain the mutual information between the two graph views. 
Specifically, the intra-bahavior contrastive loss  on the user-side is formalized as follows:
\begin{equation}
    \mathcal{L}^{b,h}_{cl_{},user}=\sum_{k=1}^{K}\sum_{u \in \mathcal{U}}-\log \frac{\exp (\phi(e^{b,k}_u, e^{h,k}_u) / \tau)}{\sum_{v \in \mathcal{U}} \exp (\phi(e^{b,k}_{u}, e^{h,k}_{v}) / \tau)},    
\label{eq:B-H_loss}
\end{equation}
where $(e^{b,k}_u, e^{h,k}_u)$ denotes the positive sample pairs  under the $k$-th behavior  and $(e^{b,k}_{u}, e^{h,k}_{v})$ is the negative sample pairs.
The calculation of  intra-bahavior contrastive loss  on the item-side ${L}^{b,h}_{cl_{},item}$ follows a similar process.
Finally, the overall intra-bahavior contrastive loss is defined as follows:
\begin{align}
    \mathcal{L}^{b,h}_{cl} = \mathcal{L}^{b,h}_{cl,user} +\mathcal{L}^{b,h}_{cl,item}.
\end{align}

\subsection{Model Prediction} 
We propose a behavior mutual enhancement module to refine current behavior by extracting and integrating information from other behaviors. For the $k$-th behavior, we update the representations of users and items as follows: 
\begin{equation}
\begin{split}
\tilde{e}^{k}_u &= \text{softmax}\left(\frac{e^k_u \cdot E^b_u}{\sqrt{d}}\right) \cdot (E^b_u)^{\mathsf{T}}, \\
\tilde{e}^{k}_i &= \text{softmax}\left(\frac{e^k_i \cdot E^b_i}{\sqrt{d}}\right) \cdot (E^b_i)^{\mathsf{T}},
\end{split}
\end{equation}
where $E^b_u = [e^1_u,e^2_u,...,e^K_u] \in \mathbb{R}^{d \times K}$,  $E^b_i = [e^1_i,e^2_i,...,e^K_i]  \in \mathbb{R}^{d \times K}$.

To obtain comprehensive representations from both global and local perspectives, we 
integrate  the $k$-th behavior-specific user/item embeddings with its corresponding global user/item embeddings as follows:
\begin{align}
    \bar{e}^{k}_u = \tilde{e}^{k}_u \oplus e^g_u,\\
    \bar{e}^{k}_i = \tilde{e}^{k}_i \oplus e^g_i.
\end{align}
Finally, the prediction score ${y}^k_{ui}$ of whether user $u$ will interact with item $i$ under behavior $k$ is defined as follows: 
\begin{equation}
    {y}^k_{ui} = ({\bar{e}^{k}_u})^{\mathsf{T}}\cdot\bar{e}^{k}_i.
\end{equation}  
  
\subsection{Joint Optimization}
In the training phase, we adopt a multi-task learning framework \cite{tang2020progressive} to explore the information within multiple behaviors in a joint manner. 
Specifically, we treat each behavior as a separate prediction task, and utilize the BPR  loss \cite{steffen2009bpr} as the optimization objective that is defined as follows: 
\begin{equation}
    \mathcal{L}_{k}=\sum_{(u, i, j) \in \mathcal{T}_k}-\ln \sigma(y_{u i}^{k}-y_{u j}^{k}),
\end{equation}  
where $\mathcal{T}_k=\{(u, i, j) \mid(u, i) \in \mathcal{R}^{+},(u, j) \in \mathcal{R}^{-}\}$ indicates the positive and negative samples in the training set, $\mathcal{R}^{+}$ ($\mathcal{R}^{-}$) denotes the sample set with observed (unobserved) interaction in the $k$-th behavior, $\sigma(\cdot)$ is the sigmoid function.

Thus, the loss function  $\mathcal{L}$ for the overall task is formulated as follows:
\begin{equation}
  \begin{aligned}
    \mathcal{L} &= \underbrace{\sum_{k=1}^K \mathcal{L}_k}_{\emph{prediction}}\!\!
    +\, \alpha ( 
    \underbrace{\lambda_1 \mathcal{L}^{g,b}_{cl} + \lambda_2 \mathcal{L}^{g,h}_{cl}}_{\emph{inter-behavior}}
    + \!\!\!\!\!\! \underbrace{ \lambda_3 \mathcal{L}^{b,h}_{cl}}_{\emph{intra-behavior}} \!\!\!\!\!\! )
    + \beta \|\Theta\|_2,
  \end{aligned}
\end{equation} 
where $\alpha$ is the coefficient of the overall consistency loss, $\beta$ is the  coefficient of the $L_2$ regularization loss, $\Theta$ are  the trainable parameters of the proposed model, and $\lambda_1$, $\lambda_2$ and $\lambda_3$ are the coefficients of three individual consistency losses, respectively.

\section{Experiments}

\subsection{Experimental Settings}

\subsubsection{{Datasets}} 
Following previous works \cite{cheng2023multi,yan2023mbhgcn,meng2023parallel}, we employ  three real-world datasets, including Beibei, Taobao and Tmall, to evaluate the performance of our proposed approach. The detailed statistics of each dataset are shown in Table \ref{tab:dataset}.

\textbf{Beibei}\footnote{https://www.beibei.com/} is widely used as the benchmark for multi-behavior recommendation. It is collected from Beibei which is the largest e-commerce platform for infant products in China. This dataset contains three types of user behaviors, i.e.,  \textit{view, cart and buy.}  

\textbf{Taobao}\footnote{https://www.taobao.com/} is collected from Taobao, which is 
one of the most popular e-commence platforms in China. It contains 15,449 users and 11,953 items with three types of user behaviors, i.e.,  \textit{view, cart and buy.}  

\textbf{Tmall}\footnote{https://www.tmall.com/} is collected from Tmall, which is also one of the most popular e-commence platforms in China. It contains  41,738  users and 11,953 items.  Unlike Beibei and Taobao, this dataset consists of four types of user behaviors, including \textit{view, collect, cart and buy}. 

\begin{table}[!th]
\begin{center}
    \caption{Statistics of the experimental datasets.}
    \begingroup
    \label{tab:dataset}
    \resizebox{1\linewidth}{!}{  
    \begin{tabular}{  c | c  c | c  c  c  c }
    \toprule
    \multirow{1}{*}{\textbf{Dataset}} & 
    \multirow{1}{*}{\textbf{Users}} & 
    \multirow{1}{*}{\textbf{Items}} &     
    \multirow{1}{*}{\textbf{View}} &     
    \multirow{1}{*}{\textbf{Collect}} &     
    \multirow{1}{*}{\textbf{Cart}} &     
    \multirow{1}{*}{\textbf{Buy}}   
    
    \\ 
    \midrule
    \textbf{Beibei}  & 21,716 & 7,977    & 2,412,586  & -       &  642,622    &  282,860  \\ 
    \textbf{Taobao}  & 15,449 & 11,953   & 873,954    & -       &  195,476    &  92,180  \\  %
    \textbf{Tmall}   & 41,738 & 11,953   & 1,813,498  & 221,514 &  1,996      &  255,586  \\  

    \bottomrule
    \end{tabular}
    }
    \endgroup
\end{center}    
\end{table}

\subsubsection{{Baselines} }
To validate the performance of our proposed approach HEC-GCN, we compare it with both single-behavior and multi-behavior recommendation methods: 

\textbf{Single-behavior recommendation methods:}
\begin{itemize} 
    \item{\textbf{MF-BPR}} \cite{steffen2009bpr}.
    It introduces a generic optimization criterion BPR-Opt for matrix factorization and leverages both positive and negative samples for optimization based on the maximum posterior estimator. 
    \item{\textbf{NeuMF}} \cite{he2017neural}. 
    This is a neural collaborative filtering based recommendation method, which utilizes both GMF and MLP to model the non-linear interactions between users and items.
    \item{\textbf{LightGCN}} \cite{he2020lightgcn}.  
    LightGCN is a lightweight GCN-based recommendation method, which attempts to model user and item representations via propagating information over the user-item bipartite graph.  
\end{itemize}

\textbf{Multi-behavior recommendation methods:} 
\begin{itemize} 
    \item{\textbf{RGCN}} \cite{michael2018modeling}. 
    RGCN is designed to handle graph learning tasks with multiple types of edges, enabling the propagation of type-specific graph information by distinguishing various types of edges.  
    \item{\textbf{GNMR}} \cite{xia2021multi}. 
    This method utilizes heterogeneous graph neural networks to model collaborative signals between different types of user-item interactions, and captures behavior dependency through recursive embedding propagation.
    \item{\textbf{NMTR}} \cite{gao2019neural}. 
     It simultaneously models user preferences under different user behaviors by using cascaded prediction, and adopts a multi-task optimization framework.
    \item{\textbf{MBGCN}} \cite{jin2020multi}. 
    MBGCN leverages  a unified multi-behavior graph to model the contribution degree of different behaviors to the target behavior, and introduces item-item graph to extract richer interaction information.
    \item \textbf{S-MBRec} \cite{gu2022self}.
    This model designs a supervised task to differentiate the importance of different behaviors while capturing commonality with a star-style contrastive learning between the target and auxiliary behaviors.
    \item{\textbf{CRGCN}} \cite{yan2023cascading-crgcn}.
    CRGCN explicitly exploits the dependencies among sequential behaviors  and designs a cascading residual graph convolutional network to gradually learn and refine user preferences. 
    \item{\textbf{MB-CGCN}} \cite{cheng2023multi}.
    It considers the behavior dependency in a behavior chain  and leverages a cascading graph convolutional network to learn user and item embeddings. The learned embeddings from each behavior are then merged for the final behavior prediction.
    \item{\textbf{MB-HGCN}} \cite{yan2023mbhgcn}. 
    This model utilizes a hierarchical graph convolutional network to learn user and item embeddings from global perspective to local behavior-specific perspective to jointly model user preferences.
    \item{\textbf{PKEF}} \cite{meng2023parallel}. 
    PKEF adopts the parallel knowledge to enhance the hierarchical information propagation to address the issue of imbalanced distribution of interactions among different behaviors. Moreover, it employs a projection mechanism to disentangle the shared and unique signals of different behaviors. 
    \item{\textbf{BCIPM}} \cite{yan2024behavior}.
    It develops a behavior-contextualized item preference network to learn item-aware preferences from user-item interactions within specific behaviors, and introduces a GCN module to capture high-order neighbor preferences of users within the target behavior. 

\end{itemize}

\subsubsection{{Implementation Details} }
The implementation of our model is based on PyTorch.  
To make a fair comparison, we follow configuration setting of  \cite{meng2023parallel}, which adopts an embedding size from  \{64, 128\}.
The learning rate and regularization coefficient are set to $5e^{-4}$ and $1e^{-3}$ respectively.
We utilize Adam \cite{diederik2015adam} to optimize the model parameters.
For the number of GCN layers in each behavior, we consider options among \{1, 2, 3\}.
Additionally, coefficients  $\lambda_1$, $\lambda_2$, and $\lambda_3$  
are adjusted in \{0, 0.5, 1.0, 1.5, 2.0, 2.5\}, with the constraint that their sum is equal to 3. 
The  coefficient $\alpha$  is tuned within \{0.1, 0.5\}.
The optimal settings for the  temperature coefficient $\tau$ and the number of hyperedges $S$ are discussed in Sect. \ref{sect:hyper}.
\subsubsection{{ Evaluation Metrics.} }
In line with previous works \cite{gao2021learning,he2020lightgcn,he2017neural}, we adopt the leave-one-out strategy for method evaluation, which implies that each user's positive sample in the test set is the last item interacted with by the user, and  the positive sample in the validation set is the penultimate interaction items.
In order to evaluate the performance of our model, we adopt Hit Ratio (HR@$n$) and Normalized Discounted Cumulative Gain (NDCG@$n$) as evaluation metrics.

\subsection{Performance Evaluation}

\begin{table*}[!t]
\begin{center}
    \captionsetup{justification=centering}
    \caption{Performance comparison on Beibei, Taobao and Tmall datasets in terms of HR@10 and NDCG@10. The best and second best scores in each column are in bold and underlined, respectively. (`Rel Impr.' indicates the relative improvement of HEC-GCN over the best performing baseline.)
    }
    \begingroup

    \resizebox{0.8\linewidth}{!}{  
    \begin{tabular}{  c | c |  c c |  c c | c c } 
    
    \toprule
    \multirow{2}{*}[-0.75ex]{{Type}} & 
    \multirow{2}{*}[-0.75ex]{{Method}} & 
    
    \multicolumn{2}{c|}{Beibei} & 
    \multicolumn{2}{c|}{Taobao} & 
    \multicolumn{2}{c}{Tmall}  \\
    \cmidrule(){3-8}
    ~  &  ~      & HR@10          & NDCG@10 & HR@10         & NDCG@10   & HR@10          & NDCG@10  \\
    \midrule
    
     \multirow{3}{*}[-0.2ex]{{Single-behavior}} & MF-BPR   & 0.0284 &  0.0142 & 0.0128 & 0.0059 & 0.0154 & 0.0072  \\ 
     ~                                                 & NeuMF    & 0.0231 &  0.0124 & 0.0077 & 0.0035 & 0.0100 & 0.0049  \\ 
     ~                                                 & LightGCN & 0.0318 &  0.0174 & 0.0381 & 0.0224 & 0.0450 & 0.0249  \\ 

     \midrule
     \multirow{3}{*}[-9ex]{{Multi-behavior}}  & RGCN     & 0.0363 &  0.0188 & 0.0215 & 0.0104 & 0.0316 & 0.0157   \\ 
     ~                                               & GNMR     & 0.0413 &  0.0221 & 0.0368 & 0.0216 & 0.0393 & 0.0193   \\ 
     ~                                               & NMTR     & 0.0429 &  0.0198 & 0.0282 & 0.0137 & 0.0536 & 0.0286   \\ 
     ~                                               & MBGCN    & 0.0470 &  0.0259 & 0.0509 & 0.0294 & 0.0549 & 0.0285   \\ 
     ~                                               & S-MBRec  & 0.0489 &  0.0253 & 0.0662 & 0.0353 & 0.0694 & 0.0362   \\ 
     ~                                               & CRGCN    & 0.0459 &  0.0324 & 0.0855 & 0.0439 & 0.0840 & 0.0442   \\ 
     ~                                               & MB-CGCN  & 0.0579 &  0.0381 & 0.1233 & 0.0677 & 0.0984 & 0.0558   \\      
     ~                                               & MB-HGCN  & 0.0744 &  0.0369 & 0.1380 & 0.0728 & 0.1470 & 0.0784  \\      
     ~                                               & PKEF     & \underline{0.1130}  &  \underline{0.0582} & 0.1385 & 0.0785 & 0.1277 & 0.0721  \\     
     ~                                               & BCIPM    & 0.0816 &  0.0406 & \underline{0.1426} & \underline{0.0807} & \underline{0.1512} &  \underline{0.0803}   \\     

    \cmidrule(){2-8}  
     ~    & \textbf{HEC-GCN} & \textbf{0.1347} &  \textbf{0.0675} & \textbf{0.1960} & \textbf{0.1061} & \textbf{0.1715} & \textbf{0.0922}  \\     
    \cmidrule(){2-8}  
     ~                                               & Rel Impr.          & 19.20\%   & 15.98\%  & 37.45\%  & 31.47\%   & 13.43\%  & 14.82\%    \\     
    
    \bottomrule
    \end{tabular}
    }
    \endgroup
    \label{tab:overall}
\end{center}   
\end{table*}

Table \ref{tab:overall} demonstrates the evaluation results of the performance comparison.  The best and second best scores in terms of each evaluation metric are in bold and underlined, respectively.  From the results, we have the following key observations: 
First, among all single-behavior methods, LightGCN is consistently superior to MF-BPR and NeuMF in all cases. The reason for the performance improvement is that LightGCN effectively captures high-order interaction information, whereas both MF-BPR and NeuMF focus solely on information from direct neighbors.
Second, among all multi-behavior methods, RGCN shows the lowest performance across all cases. This is due to it combines information from different behaviors through simple summation, which overlooks the unique characteristics of each behavior. GNMR achieves better performance than RGCN by employing heterogeneous graph neural networks, which can distinguish the unique characteristics of different behaviors. NMTR obtains a competitive performance as GNMR and also outperforms RGCN since it explores the sequential dependency between behaviors by indirectly modeling the interaction scores associated with each behavior.
    
MBGCN outperforms both GNMR and NMTR as it further captures  the contribution degree of different behaviors and assigns specific weights to them. Compared to MBGCN, S-MBRec exhibits a superior performance because it employs a supervised task to better model the importance of different behaviors. 
CRGCN and MB-CGCN achieve superior performance by leveraging cascaded graph convolutional networks to capture the cascade relationships among multiple behaviors.  
MB-HGCN yields a better performance than CRGCN and MB-CGCN. The reason is that MB-HGCN effectively leverages multi-behavior information through its hierarchical learning and aggregation strategies, 
PKEF outperforms MB-HGCN as it can learn complex interactions across multiple behaviors by integrating both cascade and parallel paradigms. 
Among all baselines, BCIPM achieves the best performance in most cases, attributed to its strong capability to capture nuanced item-specific preferences within specific behaviors using a behavior-contextualized item preference network. 

Our proposed method HEC-GCN outperforms the best performing baseline by a large margin. 
For example, the relative improvements of HEC-GCN over the best performing baseline in terms of  HR@10  on the three datasets (i.e., Beibei, Taobao and Tmall)  are 19.20\%, 37.45\% and 13.43\%, respectively. This is because HEC-GCN captures both fine- and coarse-grained hierarchical correlations by incorporating the user-item interaction graph along with a behavior-specific hypergraph. Additionally, it ensures inter- and intra-behavior consistency in representations through a contrastive learning framework.

\subsection{Ablation Study}
\begin{table*}[b]
\begin{center}
    \caption{Performance comparison of different HEC-GCN variants on all datasets.}
    \label{tab:ablation_study}
    \begingroup
    \resizebox{0.72\linewidth}{!}{  
    \begin{tabular}{ c | c c | c  c | c c}
    \toprule
    \multirow{2}{*}[-0.75ex]{Methods} & 
    \multicolumn{2}{c|}{Beibei} & 
    \multicolumn{2}{c|}{Taobao} & 
    \multicolumn{2}{c}{Tmall}  \\ 
    \cmidrule(){2-7}
    ~   & HR@10          & NDCG@10  & HR@10          & NDCG@10  & HR@10          & NDCG@10 \\
    \midrule
    HEC-GCN(full)                  & \textbf{0.1347}  &  \textbf{0.0675} & \textbf{0.1960} & \textbf{0.1061} & \textbf{0.1715} & \textbf{0.0922}  \\ 
    \midrule
    \textbf{ ${w/o \;\;global}$ }     & 0.0664           & 0.0342         & 0.1493            & 0.0861        & 0.0678            & 0.0359  \\
    \textbf{ ${w/o \;\;hyper}$ }      & 0.1220	       & 0.0616	         & 0.1699          & 0.0910         & 0.1607	        & 0.0861 \\
    \textbf{ ${w/o \;\;stop}$ }        & 0.1157           & 0.0592            & 0.1820        & 0.0978        & 0.1714            & 0.0921  \\
    \textbf{ ${w/o \;\;cascading}$ }   &  0.1243           & 0.0630            & 0.1766        & 0.0949        & 0.1633            & 0.0873  \\
    \textbf{ ${w/o \;\;mutual}$ }      &  0.1076          & 0.0550            &  0.1768       & 0.0946        & 0.1585            & 0.0860  \\
     
    \bottomrule
    \end{tabular}
    }
    \endgroup
\end{center}
\end{table*}

In this section, we conduct  ablation study to  elaborate the reasons for the performance improvement and  verify the rationality of the designed key components in our model.  
The variants of our model are designed as follows:
\begin{itemize} 
\item ${w/o \;\;global}$: We remove the global graph  as well as the cross-behavior contrastive loss associated with it. 
\item ${w/o \;\;hyper}$: We discard behavior-specific hypergraphs from the hypergraph enhanced cascading graph convolution network module.
\item ${w/o \;\;stop}$:  The stop-gradient strategy  between the interaction graph  and the hypergraph  across behaviors is removed.
\item ${w/o \;\;cascading}$: The cascading architecture is eliminated, and embeddings for multiple behaviors are all initialized based on the global graph.
\item ${w/o \;\;mutual}$: We remove the behavior mutual enhancement module, and directly adopt embeddings learned from each behavior for prediction.
\end{itemize}
The results are shown in Table \ref{tab:ablation_study}. 
We can observe that each component contributes significantly to enhancing the performance of our proposed approach. To be specific, we observe that:  
First, removing the global graph (${w/o \;\;global}$) leads to a significant performance degradation. This is because it plays a vital role in learning the user's overall preferences across various behaviors. Moreover, it acts as a mediator for cross-behavior contrastive learning, which imposes an inherent constraint to maintain consistency across different behaviors. 
Second, the removal of hypergraph  (${w/o \;\;hyper}$) results in a considerable decline in model performance, highlighting the importance of incorporating learnable hypergraphs. This result is attributed to the hypergraph’s ability to capture implicit high-order correlations among users or items, thereby complementing the behavior interaction graph.
Third, discarding the stop-gradient strategy causes a performance drop of the proposed model,  which reveals the importance of maintaining the unidirectional information propagation from fine-grained interaction graph learning to  coarse-grained behavior-specific hypergraph learning.   
Fourth,  the results of the variant ${w/o \;\;cascading}$ validate the benefits of modeling the dependencies among sequential behaviors, which is consistent with previous works. 
Finally, the variant ${w/o \;\;mutual}$,  which excludes behavior mutual enhancement, shows a significant performance decline, highlighting the effectiveness of integrating information from other behaviors.

\subsection{Effectiveness Analysis of Behavior Consistency-Guided Contrastive Learning }
\begin{table*}[b]
\begin{center}
    \caption{The impact of contrastive learning constraints on model performance.}
    \label{tab:ablation_study_ssl}
    \begingroup
    \resizebox{0.72\linewidth}{!}{  
    \begin{tabular}{ c | c c | c  c | c c}
    \toprule
    \multirow{2}{*}[-0.75ex]{Methods} & 
    \multicolumn{2}{c|}{Beibei} & 
    \multicolumn{2}{c|}{Taobao} & 
    \multicolumn{2}{c}{Tmall}  \\ 
    \cmidrule(){2-7}
    ~   & HR@10          & NDCG@10  & HR@10          & NDCG@10  & HR@10          & NDCG@10 \\
    \midrule
    HEC-GCN(full)               & \textbf{0.1347}  &  \textbf{0.0675} &  \textbf{0.1960}  &  \textbf{0.1061}  &  \textbf{0.1715} & \textbf{0.0922}  \\ 
    \midrule
    \textbf{  $w/o \;\;CL_{intra}$ }                      & 0.1200	       & 0.0605          & 0.1886          & 0.1013         & 0.1681        & 0.0902 \\
    \textbf{$w/o \;\;CL_{cross}$ }                     & 0.0970           & 0.0498          & 0.1671          & 0.0877	        & 0.1531        & 0.0794 \\
    \textbf{ $w/o \;\; CL_{all}$ }          & 0.0854           & 0.0438          & 0.1343	       & 0.0709         & 0.1400        & 0.0745 \\
    \bottomrule
    \end{tabular}
    }
    \endgroup
\end{center}
\end{table*}

To investigate the contribution of the inter- and intra-behavior consistency constraints based on contrastive learning, we conduct experiments by removing the constraints from HEC-GCN and obtain the following variants:  
\begin{itemize}
\item $w/o \;CL_{intra}$: We discard the consistency constraint between each user-item
interaction graph and its corresponding hypergraph in the same behavior, i.e., the intra-behavior contrastive loss $\mathcal{L}^{b,h}_{cl}$ is removed.
\item $w/o \;CL_{cross}$: We remove the consistency constraint across different behaviors, i.e., the inter-behavior contrastive loss ($\mathcal{L}^{g,b}_{cl}$ and $\mathcal{L}^{g,h}_{cl}$) is ignored. 
\item $w/o \;CL_{all}$: Both intra- and inter-behavior contrastive losses are simultaneously removed.
\end{itemize}

The results are exhibited in Table \ref{tab:ablation_study_ssl}. We observe that both intra- and inter-behavior contrastive constraints play a crucial role in HEC-GCN, and removing the inter-behavior contrastive constraint results in a greater decline in model performance than removing the intra-behavior contrastive constraint on all datasets. Furthermore, discarding both constraints leads to an even more severe performance degradation, indicating that the intra- and inter-behavior contrastive constraints can complement each other.



\begin{figure}[!t]
\centering
\includegraphics[width=\linewidth]{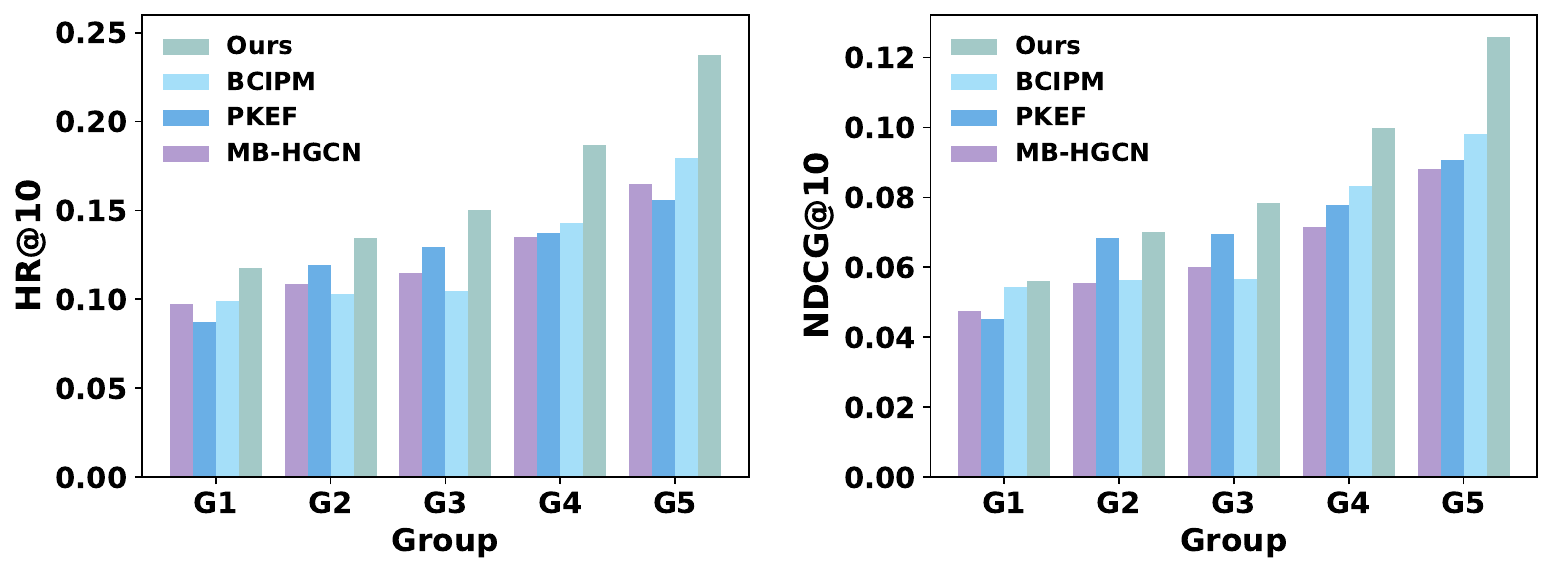}
\caption{  
Model performance with respect to different interaction density degrees.
} 
\label{fig:fig_buck}
\end{figure}

\begin{figure}[!t]
\centering
\includegraphics[width=\linewidth]{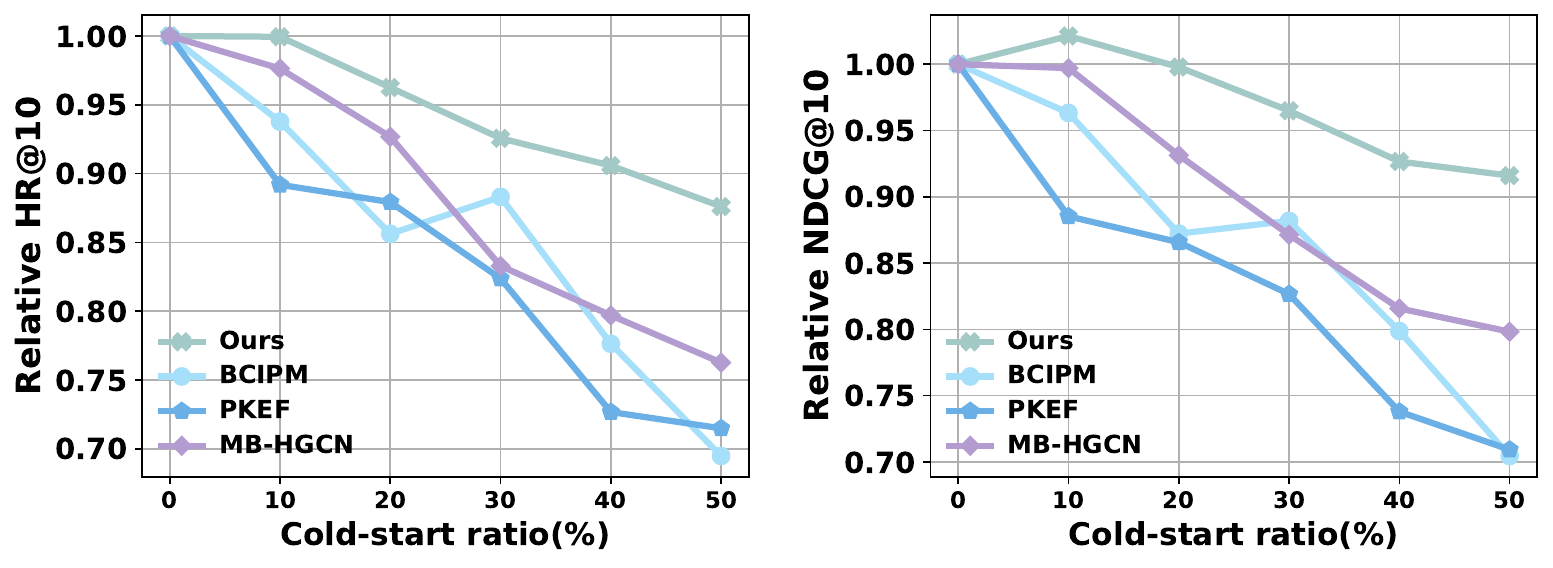}
\caption{ 
The impact of different cold-start ratios on model performance.
}
\label{fig:fig_cold}
\end{figure}
 

\subsection{Performance  with Different Interaction Density Degrees.} 
To assess the performance of our method across different levels of interaction density, we divide users in the test set into five categories based on their number of interactions in the Taobao dataset. Specifically, we separate users into five categories: G1 (1-3 interactions), G2 (4-6 interactions), G3 (7-9 interactions), G4 (10-12 interactions) and G5 (over 12 interactions). 
We then report the performance of our proposed model for each category in Fig. \ref{fig:fig_buck}. 

Compared to the three state-of-the-art baselines (e.g., MB-HGCN, PKEF and BCIPM), our method consistently demonstrates the optimal recommendation performance across all categories. 
Furthermore, the relative performance improvement of our model over these strong baselines grows as the interaction density levels increase.
The results verify the effectiveness of incorporating learnable hypergraphs combined with intra- and inter-behavior contrastive constraints in our proposed approach. 

\subsection{Performance in Cold-start Scenario} 
In this section, we investigate the effectiveness of HEC-GCN against the cold-start issue. 
To evaluate the influence of cold-start degrees on the performance our proposed model, we randomly sample 50\% of the users in Taobao dataset and remove their historical interactions proportionally, i.e., 10\%, 20\%, 30\%, 40\%, 50\% of interactions are discarded in our experiments. We compare  HEC-GCN  with MB-HGCN, PKEF and BCIPM. 
Fig. \ref{fig:fig_cold} shows the relative performance degradation of different methods in terms of HR@10 compared to the performance on full interaction data. Our approach demonstrates less performance degradation compared to all baselines. This is primarily due to the following reasons: 1) The behavior-specific hypergraph in HEC-GCN is effective in capturing complex high-order correlations among users and items, which complements its corresponding behavior-specific interaction graph to mitigate the cold-start issue. 2) HEC-GCN incorporates the inter-behavior consistency constraint which can further strengthen the representation learning in each behavior by incorporating information from other behaviors.

\subsection{Impact of Auxiliary Behaviors}
\begin{figure}[!t]
\centering
\includegraphics[width=\linewidth]{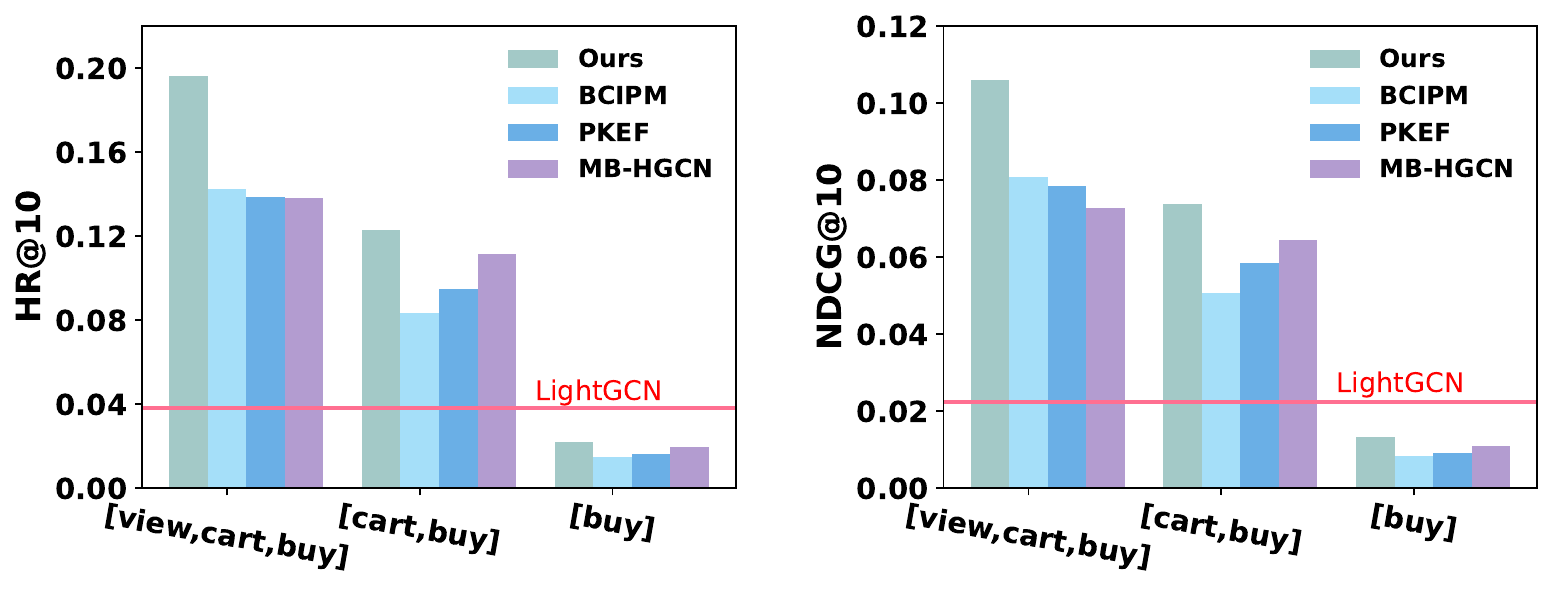}
\caption{  
The Impact of auxiliary behaviors on performance.
}
\label{fig:fig_bo}
\end{figure}

To examine how auxiliary behaviors affect model performance, we gradually remove these behaviors and assess the performance of our proposed method, HEC-GCN, alongside three competitive baselines:  MB-HGCN, PKEF and BCIPM. Fig \ref{fig:fig_bo} illustrates the results on the Taobao dataset. 

We observe that removing the auxiliary behavior ``view'' leads to a significant performance decline of all methods. In addition, a further performance drop can be observed if we continue to discard the    auxiliary behavior ``cart'', i.e., we only leverage the target behavior for recommendation. 
The results demonstrate the positive impact of incorporating more auxiliary behaviors across all methods, with our proposed approach making the most effective use of these behaviors.
Moreover, among different settings of auxiliary behaviors, HEC-GCN consistently outperforms all three strong baselines, which suggests the better capability of HEC-GCN for effectively exploring multi-behavior information.
Additionally, when only utilizing the target behavior ``buy'', the performance of all multi-behavior recommendation models, including our proposed HEC-GCN, is lower than that of a single-behavior recommendation model like LightGCN \cite{he2020lightgcn}. This can be attributed to the complex structure of these multi-behavior models, which may be more prone to overfitting.

\subsection{Impact of Behavior Chain Order}
\label{exp:impact_bo}

\begin{figure}[!b]
\centering
\includegraphics[width=\linewidth]{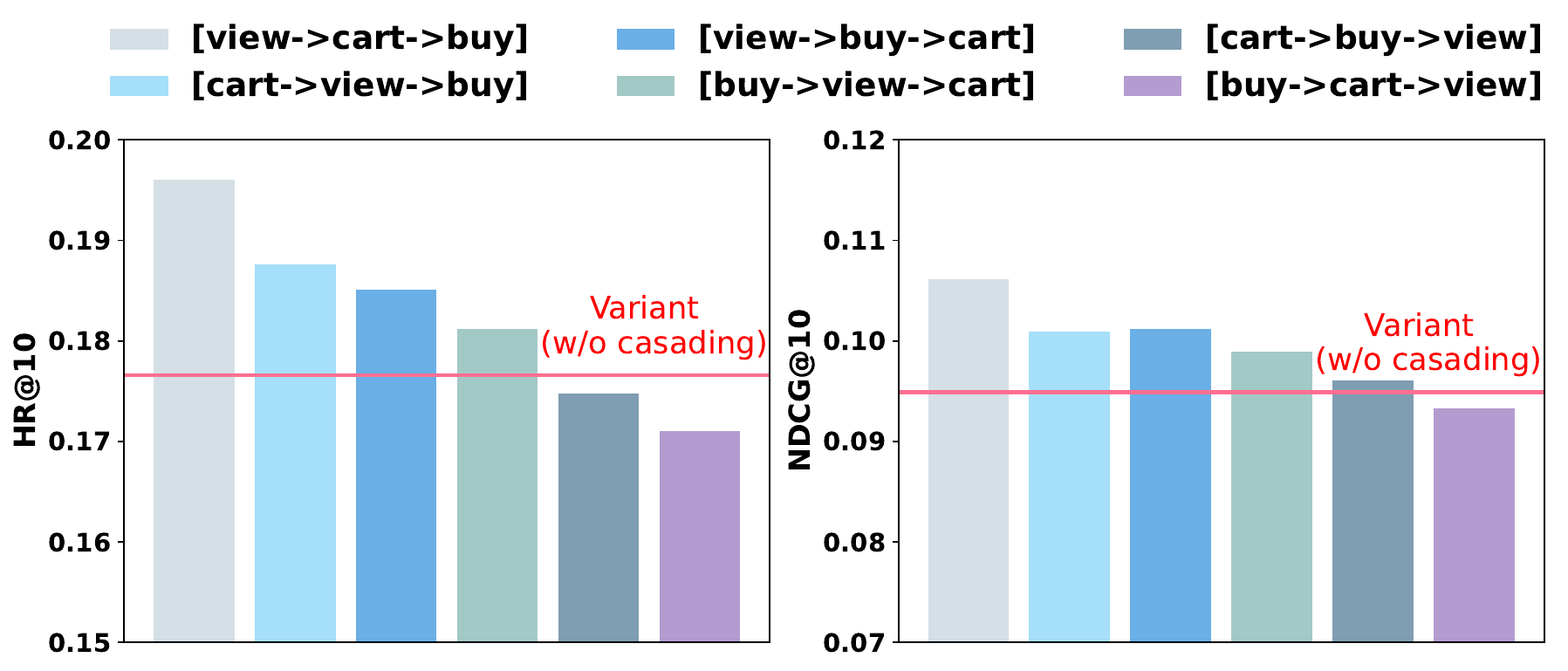}
\caption{ 
Impact of behavior sequence order on performance in the cascading architecture.
}
\label{fig:fig_aux_bo}
\end{figure}

In a behavior chain, a later behavior generally offers a stronger indication of user preference than an earlier one \cite{cheng2023multi}, where we refer to the earlier behavior as  causal behavior. 
To investigate the effect of behavior  order in the chain on the performance of HEC-GCN, we adjust the order of behaviors on the Taobao dataset, and the results are shown in Fig \ref{fig:fig_aux_bo}. 

We can  observe that the original behavior chain [view $\rightarrow$ cart $\rightarrow$ buy] results in the best  performance, whereas its reverse order yields the worst performance, even worse than the variant without a cascading architecture (i.e., variant $w/o\;\;cascading$).  
In addition, placing more casual behaviors later in a behavior china leads to a performance decline  of our proposed model. For example, the model performance on the behavior chain [buy $\rightarrow$ view $\rightarrow$ cart] is inferior to that on the behavior chain [view $\rightarrow$ buy $\rightarrow$ cart].
This result indicates that the sequence of behaviors plays a crucial role in the effectiveness of our proposed method. The main reason is that casual behaviors, such as ``view'', often contain more noise. By positioning these casual behaviors earlier in the sequence, the cascading architecture can progressively refine the learning of user preferences, allowing a more effective transfer of user preferences from auxiliary behavior to the target behavior.


\subsection{Hyperparameter Analysis}
\label{sect:hyper}
\begin{figure}[!t]
\centering
\includegraphics[width=\linewidth]{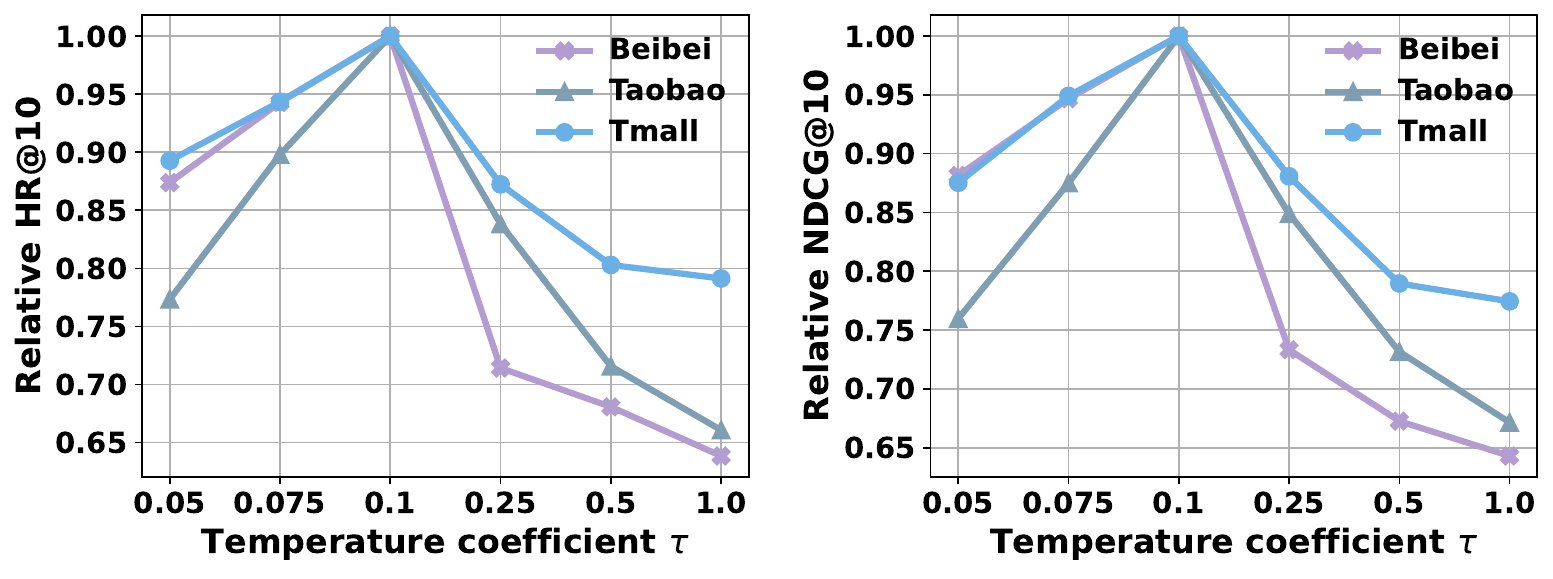}
\caption{The Impact of temperature coefficient $\tau$.
}
\label{fig:tau_}
\end{figure}

\begin{figure}[!t]
\centering
\includegraphics[width=\linewidth]{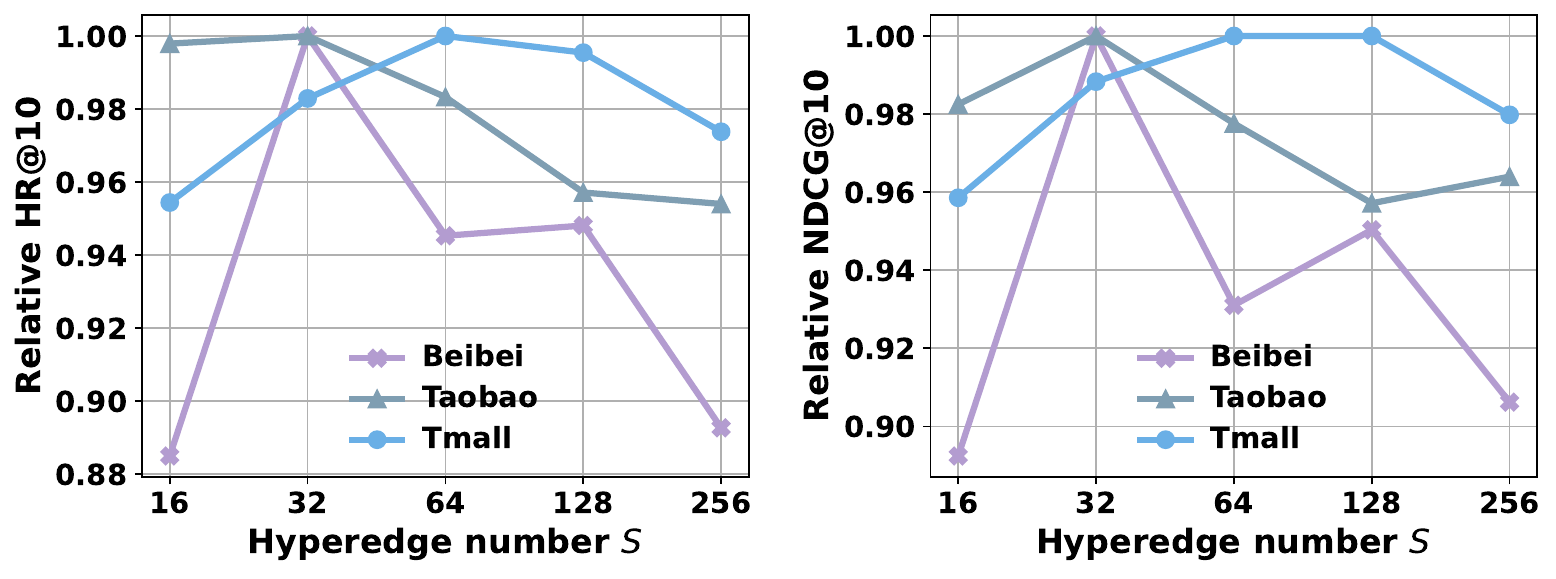}
\caption{The Impact of hyperedge number $S$.
}
\label{fig:hyper_}%
\end{figure}

In this section, we first explore the effect of the temperature coefficient $\tau$ on the model performance. Then, we analyze the influence of the number of hyperedges $S$. The changing trends of the two parameters are shown in Fig. \ref{fig:tau_} and Fig. \ref{fig:hyper_}, respectively.

The parameter $\tau$ controls the smoothness of embedding similarity, i.e., a small $\tau$ results in sharper similarity, while a larger $\tau$ produces a smoother similarity. 
We tune $\tau$ from the set of values \{0.05, 0.075, 0.1, 0.25, 0.5, 1.0\} and the corresponding results on three datasets are demonstrated in Fig. \ref{fig:tau_}. 
We can observe that,  with the increase of the value of $\tau$, the model performance  first gradually rises until it reaches the peak at $\tau$=0.1,  and it starts to decrease with a larger value of $\tau$.  

The parameter $S$ indicates the number of hyperedges utilized in a behavior-specific hypergraph. 
Each hyperedge serves as a mediator to connect users (items), and  reflects a specific 
semantic dependency between users (items). We study the effect of the number of hyperedges on model performance. To be specific, we vary the value of $S$ in \{16, 32, 64, 128, 256\}, and the results are shown in Fig. \ref{fig:hyper_}. We observe that when $S$ is small, HEC-GCN performs poorly. This could be due to the insufficient number of hyperedges, which limits the model’s ability to effectively capture the complex semantic dependencies among users (items). As $S$  gets larger, the model performance improves progressively.  If we continue to raise the value of $S$, the performance will start to drop.
The reason is that when $S$ becomes too large, it may suffer from over-fitting issue and harm model performance. The optimal performance is achieved by setting a relatively small value of $S$.

\section{Conclusion and Future Work}
In this paper, we present a novel model HEC-GCN for multi-behavior recommendation. HEC-GCN examines both fine- and coarse-grained correlations among users or items for each behavior by jointly modeling the behavior-specific interaction graph and its corresponding hypergraph, where the latter is designed to capture coarse-grained correlations among users or items, helping to mitigate the sparsity issues inherent in the interaction graph. 
Additionally, it develops a behavior consistency-guided alignment strategy that ensures representation consistency between the interaction graph and its corresponding hypergraph within each behavior, while also maintaining consistency across different behaviors.
Extensive experiments on three real-world benchmark datasets demonstrate that our proposed method considerably outperforms state-of-the-art baseline methods. 

%
%
%

\printcredits

\section*{Acknowledgement}
This work was supported by the National Natural Science Foundation of China [grant number 62472059,62372059]; the Natural Science Foundation of Chongqing, China [grant number CSTB2022NSCQ-MSX1672]; the Chongqing Talent Plan Project, China [grant number CSTC2024YCJH-BGZXM0022]; the Major Project of Science and Technology Research Program of Chongqing Education Commission of China [grant number KJZD-M202201102].

\bibliographystyle{cas-model2-names}
\bibliography{ref}
\end{document}